\documentclass[12pt]{iopart}
\usepackage{iopams}
\usepackage{graphicx} 
\usepackage{amsmath} 
\usepackage{amssymb}
\usepackage[left, pagewise]{lineno}
\usepackage[dvipsnames]{xcolor}

\newcommand{\dpsidxi}{\frac{d \psi_t}{d \xi}}
\newcommand{\dthetadxi}{\frac{d \theta}{d \xi}}

\newcommand{\xx}{\mathbf{x}}
\newcommand{\vv}{\mathbf{v}}
\newcommand{\dd}{\mathbf{d}}
\newcommand{\BBB}{\mathbf{B}}
\newcommand{\RR}{\mathcal{R}}
\newcommand{\SL}{\mathrm{SL}_2(\mathbb{R})}
\newcommand{\Trr}{\mathrm{Tr}} 
\newcommand{\Det}{\mathrm{Det}}
\newcommand{\MM}{\mathsf{M}}
\usepackage[normalem]{ulem}

\begin{document}
\title{The topology of non-resonant stellarator divertors}
\author{Robert Davies${}^1$, Christopher B. Smiet${}^2$, 
Alkesh Punjabi${}^{3}$, Allen H. Boozer${}^4$, 
Sophia A. Henneberg${}^1$}
\address{${}^1$Max Planck Institute for Plasma Physics, Wendelsteinstraße 1, 17491 Greifswald,
Germany \\
${}^2$Ecole Polytechnique Fédérale de Lausanne (EPFL), Swiss Plasma Center (SPC), CH-1015 Lausanne, Switzerland  \\
${}^3$Department of Mathematics, Hampton University, Hampton, VA 23668, USA \\
${}^4$Columbia University, New York, New York 10027, USA
}
\date{\today}

\begin{abstract}
We apply topological methods to better understand how the magnetic field in the stellarator edge can be diverted away from the confined region. Our primary method is calculating the winding numbers of closed contours, which gives information on the number and nature of fixed points within a bounded region. We first apply this to the non-resonant divertor (NRD) Hamiltonian system, and present a simple explanation for the system's diversion: trajectories are guided away from the confined region by X-points which are ``unpaired" i.e. do not have corresponding O-points and therefore do not resemble an island chain. We show how similar phenomena can occur in a similar, axisymmetric Hamiltonian system. 
Secondly, we find examples of neoclassically optimised stellarators in the QUASR database which divert the magnetic field via unpaired X-points. We present and discuss three examples, each containing novel phenomena which might be desirable for stellarator divertors. These findings broaden the horizons of how magnetic fields can be diverted in realistic stellarators, and may be attractive for future experiments and stellarator reactor design. 
\end{abstract}

\maketitle

\section{Introduction}\label{sec:intro}
\begin{figure}
    \centering
    \includegraphics[width=0.6\linewidth]{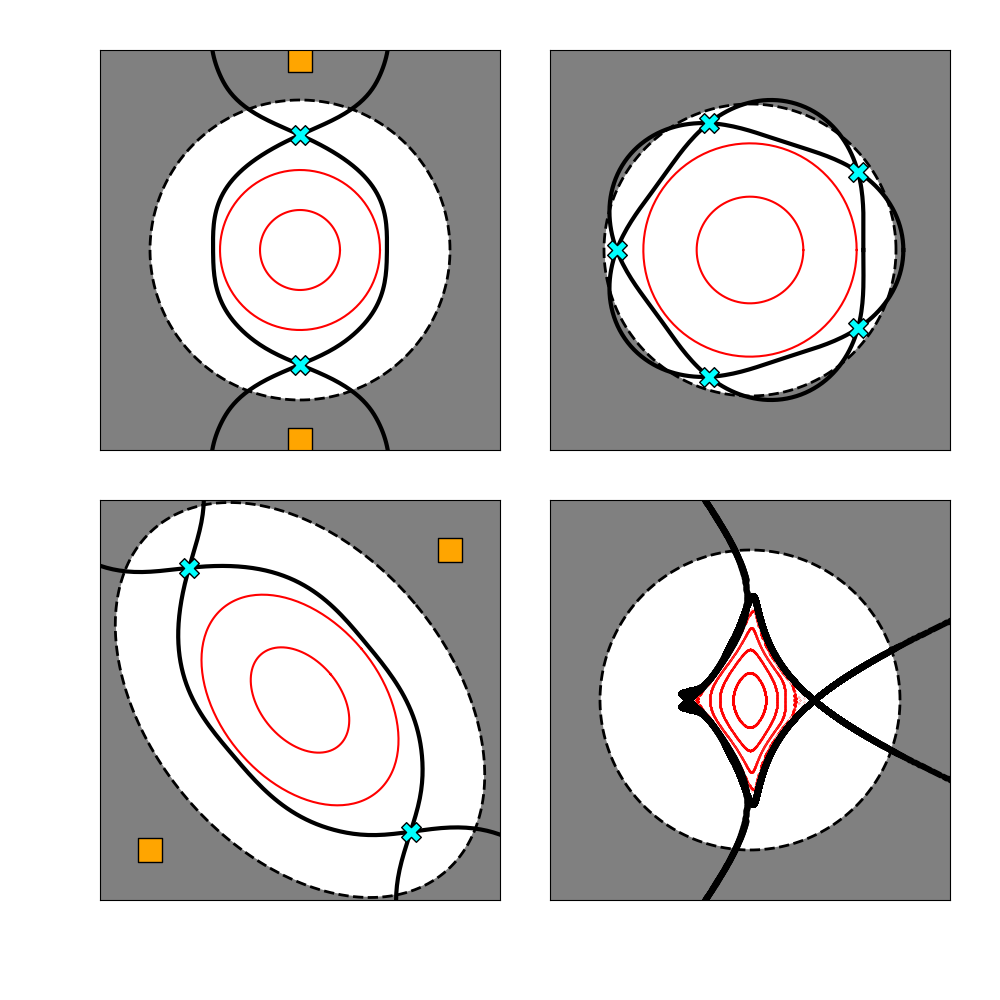}
    \caption{Schematic illustration of the most popular methods of diversion in magnetically confined fusion plasmas. In each case, nested flux surfaces of the confined plasma are shown in red and solid components are shaded grey, with the plasma-facing components (PFCs) outlined in dashed black. Upper left: A tokamak double null configuration. External coils (shown as orange squares) generate X-points (cyan crosses). This forms a separatrix (solid black line) which diverts plasma to the PFCs. Upper right: island divertor of a stellarator. The separatrix of the island chain (solid black line) intersects the PFCs. Lower left: a helical divertor. Helically rotating X-points (cyan crosses) with a helically rotating separatrix (black lines) are generated by helically rotating coils (orange squares). Lower right: the non-resonant divertor Hamiltonian, which is the focus of this work. Field lines beyond the confined plasma travel to the wall in collimated bundles (solid black lines).}
    \label{fig:cartoon}
\end{figure}
Magnetically confined fusion devices require an exhaust solution, that is, the heat and particles escaping the confined region must be managed to meet certain requirements \cite{soboleva1997energy, stangeby2000tutorial, boozer2015stellarator}. A prominent feature of most exhaust solutions is using the magnetic field to divert the escaping plasma away from the confined region. The physical separation between the confined plasma and the solid plasma-facing components (PFCs) allows plasma temperatures at the plasma-surface interface to be reduced. It also makes it harder for neutral particles and ions to re-enter the confined plasma; this is desirable for Helium exhaust and preventing PFC-generated impurities (sputtered material, for example) from diluting and polluting the fusion plasma. It is therefore highly relevant to the fusion effort to understand how magnetic fields can ``divert".

The focus of this paper is the topology of diverting magnetic fields, although this provides only partial answers to the many interrelated factors which determine the efficacy of an exhaust solution. These factors include: (1) the location, size and shape of the plasma-surface interface region; (2) the fraction of the power which can be radiated, which may be aided by sowing impurities into the edge plasma; (3) the transport of plasma heat and density along and across magnetic field lines; (4) the PFC-to-PFC connection length\footnote{Connection length is defined as the distance traveled along a magnetic field line from one PFC to another i.e. the length of a magnetic field line within the vacuum vessel/PFC-enclosed region.} (which directly impacts factors 1-3) and (5) the controllability of the edge conditions as the confined plasma evolves. These factors depend upon not only on the magnetic field structure, but also on the design of the PFCs and how the machine is operated. Nevertheless, understanding divertor topologies allows one to gain a physical understanding of how these factors might be adjusted and optimised.

Tokamaks and stellarators typically depend upon a particular topological feature, the X-point (described in section \ref{sec:fixed_points}), for diversion. Tokamaks generate their X-points using a ``separatrix coil" carrying a current parallel to the plasma current \cite{soboleva1997energy, stangeby2000tutorial} (see figure \ref{fig:cartoon}, upper left, for a schematic illustration). The poloidal magnetic field becomes zero and the X-point forms where the plasma current field and the separatrix coil field cancel each other out. At this location, the resulting rotational transform, denoted $\iota$, is also zero, since $\iota$ represents the average ratio of poloidal to toroidal turns of the magnetic field lines. Magnetic field lines which come arbitrarily close to the X-point define a separatrix, which diverts plasma onto the PFCs.

For stellarators, the most prominent divertor solutions are the island divertor as exhibited in W7-X \cite{renner2004}, the helical divertor used in LHD \cite{ohyabu1992helical, masuzaki2006overview} and the non-resonant divertor (NRD) (discussed shortly). These are schematically shown in figure \ref{fig:cartoon}. The island divertor diverts plasma via a chain of magnetic islands, containing X-points and an equal number of O-points. The X- and O-points are non-axisymmetric and their rotational transform $\iota$ is a low-order rational number. Analogously to the tokamak, the separatrix of the island chain diverts plasma and intersects the PFCs. LHD features helically rotating X-points generated by a pair of helically rotating magnetic coils. These X-points, which are embedded in a region of chaotic magnetic field, guide plasma away from the confined region onto divertor plates. Thus, for both island and helical divertors, ``diversion" of the magnetic field can be understood by studying a small number of X-points and the  magnetic field lines which pass near to these X-points. These field lines carry plasma outwards from the reactor, and the PFCs are designed to intersect these channels. The geometry of the field can be understood topologically, but in the design of a divertor careful consideration must be taken to place the wall such that the outgoing plasma is properly intercepted. Understanding the non-resonant divertor from a topological perspective is the focus of this work.

The term ``non-resonant divertor" was coined in 2015 by Boozer \cite{boozer2015stellarator}. The definition emphasised (1) the relationship between sharp curvature of flux surfaces and X-points in the magnetic field and (2) the importance of making perturbations to the magnetic field which are not resonant with the rotational transform of the field, and therefore do not produce an island divertor. Following this definition and exploiting the fact that stellarator magnetic fields are (usually) analogous to Hamiltonian systems (see section \ref{sec:fixed_points}), the NRD concept has since been investigated using a simple Hamiltonian system \cite{boozer2018simulation, punjabi2020simulation, punjabi2022magnetic, boozer2023magnetic} (described in section \ref{sec:pb_nrd}). This exhibits good diversion, in the sense that trajectories started near the last good magnetic surface\footnote{The "last good magnetic surface" in this paper means the outermost toroidally nested magnetic surface. A magnetic surface (i.e. flux surface) is a closed surface for which the magnetic field line at any point on the surface remains on the surface.} (LGMS) are carried away from the confined region in a small number of collimated bundles which are usually labeled ``turnstiles".

Accompanying these investigations, the term ``non-resonant" has also been applied to the stellarator experiments HSX \cite{bader2017hsx, garcia2024resilient} and CTH \cite{garcia2023exploration, allen2024studies}. These studies stress the following features: (1) the strike locations on the PFCs (that is, where the diverted trajectories strike the vessel wall) in poloidal and toroidal space coincide with where the LGMS is strongly curved; (2) strike locations are resilient to changes in the magnetic field, for example, changes in the $\iota$ profile; and (3) ``tendrils" of long connection length are present which correlate with strike locations. Qualitative similarities have been found between these systems and the NRD Hamiltonian, leading to a growing confidence that the underlying dynamics are the same. The connection between LGMS shaping and resilient strike locations was made, for Helias stellarators, several decades earlier \cite{strumberger1992magnetic, strumberger1992topology, nuhrenberg1992development}, although this predates the term ``non-resonant divertor".

The resilient diversion reported by NRD studies have led to a growing desire to incorporate the NRD into the design of future experiments \cite{parra2024flexible, pablant2024compelling} and commercial reactors \cite{tang2024divertor}. This enthusiasm behooves us to better understand the features of the NRD Hamiltonian system; for example, understanding the number, location and nature of the ``turnstiles". That is the goal of this paper, in which we show that $\iota=0$ X-points define the location of the outgoing trajectories in the NRD Hamiltonian system. These X-points are ``unpaired" i.e. do not have corresponding O-points (such as occurs in an island chain), causing the diverted trajectories to make arbitrarily large excursions from the confined region. 

This discovery motivates the second part of this paper, a scan over optimised stellarator configurations to find other examples of unpaired X-points. By finding such examples, we show that unpaired X-points are a general feature of stellarators, and thus could provide a means for reliable diversion. We also show examples of exotic edge topologies, which expand the horizons of possibility for stellarator divertors and are thus of great relevance to reactor design.

Therefore, this work addresses two questions: (1) Can the location of outgoing trajectories in the NRD Hamiltonian system be understood as the behaviour of ``unpaired” X-points? and (2) Do unpaired X-points typically arise in stellarators? This paper is organised as follows. We first (section \ref{sec:fixed_points}) present a short theoretical introduction to magnetic topology and topological methods. These ideas are then applied in section \ref{sec:pb_nrd} to the NRD Hamiltonian system to answer question (1). Question (2) is addressed in section \ref{sec:quasr_db}, in which we apply topological calculations to the QUASR database of stellarators. Finally, we present a conclusion and outlook (section \ref{sec:conclusions}).

\section{Magnetic topology in toroidal devices}\label{sec:fixed_points}
A useful starting point for discussing the topology of the field lines is to utilise the Poincar\'e map. The map is defined by the intersection of a trajectory in a dynamical system with a lower-dimensional subspace (in our case, the intersection of a 3D magnetic field line with a 2D plane). In fusion devices we call this map the field line map and define it by integrating the magnetic field lines over the toroidal angle $\phi$ from a point $\xx=(R,Z)$ at 
$\phi = \phi_0$ to $\phi=\phi_0+N* \tfrac{2\pi}{n_\text{fp}}$. 
Here we use cylindrical coordinates $(R, \phi, Z)$, $N$ is an integer and $n_\text{fp}$ is the number of field periods if the field possesses internal symmetries, otherwise $n_\text{fp}=1$. 
A Poincar\'e plot is generated by iterating this map multiple times, and illustrates the long-term behavior of trajectories. 

A fixed point of a map $f(\xx)$ is a point where $f(\xx)=\xx$, i.e. a point in the section where the field line passing through it returns exactly to where it started.
The magnetic axis is a well-known example. X- and O-points of an island chain are fixed points of the Poincar\'e map for some specific value of $N$. 

\begin{figure}[h]
    \centering
    \includegraphics[width=0.5\linewidth]{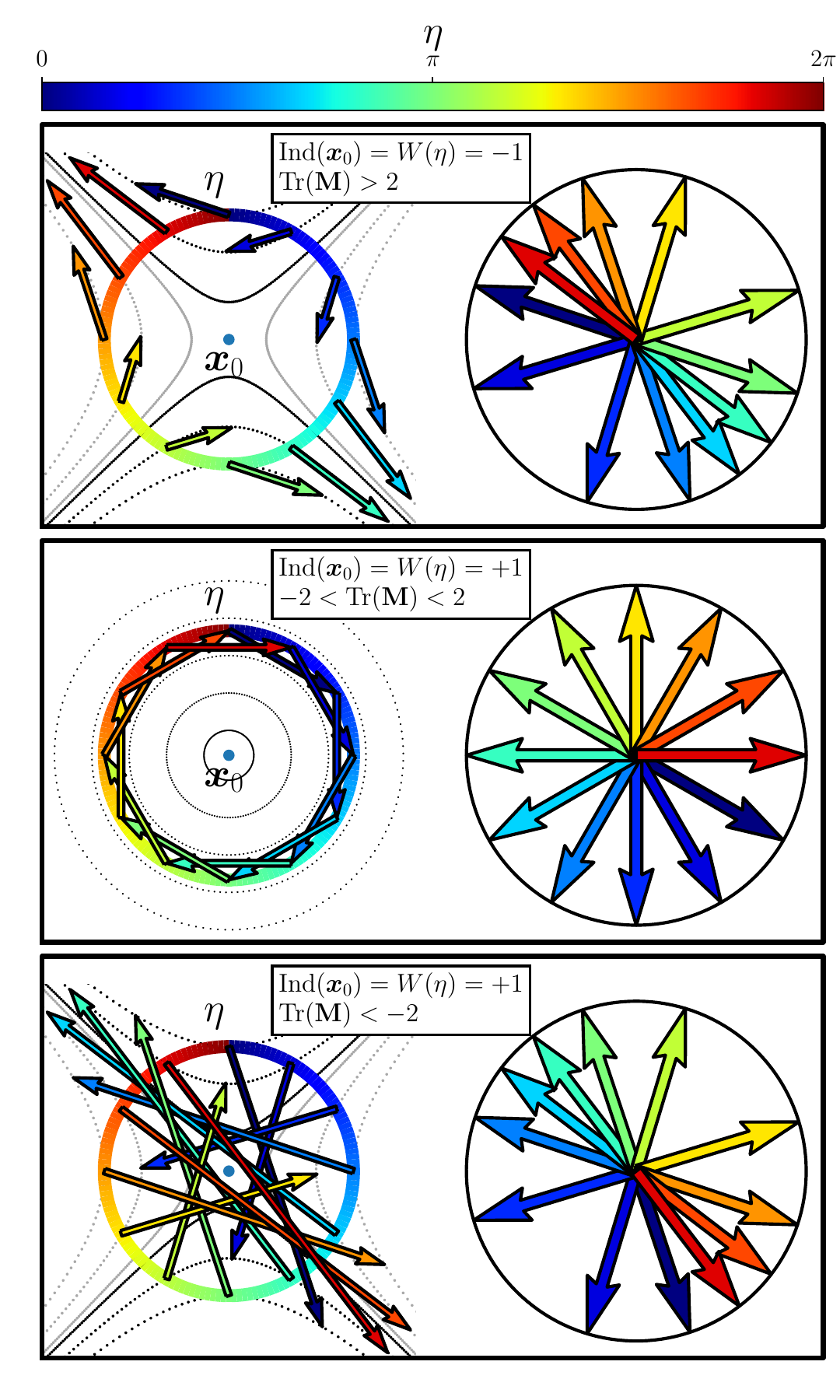}
    \caption{Illustration of the definition of the topological index of a singular fixed point in two dimensions. The topmost row corresponds to an X-point, the middle row to an O-point and the bottommost row to a hyper-hyperbolic fixed point. The left subplots show the fixed point $\xx_0$ (blue dot), surrounded by a closed curve parametrized by $\eta$, that is traversed in clockwise direction (from blue to green to red). The displacement arrows are shown for each point on this curve in colors corresponding to $\eta$ (from blue at $\eta=0$ to red at $\eta=2\pi$). When these vectors are given unit length, they map to points on the unit circle (shown in the right subplots). The index of the enclosed fixed point is the degree of the mapping from $\eta$ to the unit circle (the oriented number of times the circle is covered). For the X-point (topmost row), the topological index is $-1$, because the vectors in the right subplot rotate anticlockwise as $\eta$ increases. The topological index of the lower two rows is $+1$ because the vectors in the right subplots rotate clockwise.}
    \label{fig:indexcalc}
\end{figure}

Fixed points have a property called their topological index~\cite{poincare1928oeuvres, greene1968two}, which we study extensively in sections \ref{sec:pb_nrd} and \ref{sec:quasr_db}. The topological index can be understood as follows.
Take a closed curve $\eta$ that encloses one singular fixed point. 
Consider the set of vectors connecting each point of the original curve to its image (i.e the position after applying the Poincar\'e map), and give them unit length so they lie on the unit circle. 
These vectors vary smoothly as one traverses the original curve, and the number of full rotations they make in the direction that the original curve is traversed is the winding number $W(\eta)$ of this curve, and the topological index of the fixed point. 
An identical definition is that the index is the \emph{degree} of the map from the curve $\eta$ to the unit circle. 
See figure~\ref{fig:indexcalc} for an illustration of this, or~\cite{arnold1992ordinary_index} for a more in-depth accessible explanation. 

It can be shown that the winding number over an arbitrary closed curve $\eta$ is equal to the sum of topological indices of the fixed points within $\eta$. We use this principle extensively in this work; by calculating the winding number over a contour $\eta$, one knows the difference between the number of positive- and negative-index fixed points within the contour. 

\subsection{The Jacobian of fixed points and its properties}\label{sec:jacobian}
The nature of a fixed point can be understood by examining the Poincar\'e map in its vicinity. To first order, this is described by the Jacobian $\mathsf{M}$ of the fixed point: 
\begin{equation}\label{eq:linearmap}
    f(\xx+\delta\xx)\simeq f(\xx) + \MM \delta\xx ,
\end{equation}
where $\xx$ is the location of the fixed point, $\delta \xx$ is a small displacement and $\MM_{ij}=\partial_j f_{f,i}$ is the matrix of partial derivatives. 
This matrix is heavily constrained by the divergence-free nature of magnetic fields ($\nabla \cdot \BBB = 0$), translating to the constraint that the determinant of $\MM$ equals one ($\mathrm{Det}(\MM)=1$) \cite{solov1970plasma}. 

Hence, $\MM$ is a $2\times2$ matrix with real coefficients and a unit determinant. 
The group of such matrices is called the special linear group $\SL$. The properties of the fixed point are determined by the trace of $\MM$, the eigenvalues of $\MM$ and the eigenvectors of $\MM$ (denoted $\Trr(\MM)$, $\lambda_i$ and $\vv_i$ respectively).

The trace determines the subset\footnote{note: not sub\emph{groups}} of $\SL$ to which $\MM$ belongs. There are three options, each acting as a different area-preserving transformation on the plane:
\begin{enumerate}
    \item Elements of the elliptic subset $\{\MM\in\SL : |\Trr(\MM)|<2\}$ act as rotations of the plane. 
    \item Elements of the parabolic subset $\{\MM\in\SL: |\Trr(\MM)|=2\}$ contain the ($\pm$) identity, and elements that shear the plane, leaving one line of points fixed. 
    \item Elements of the hyperbolic subset $\{\MM\in\SL : |\Trr(\MM)|>2\}$ squeeze the plane in one direction, and stretch it in another. 
\end{enumerate}

In the context of tokamak/stellarator magnetic fields, these correspond to: 
(1) O-points such as the magnetic axis and the center of islands, where nearby field lines map to invariant circles enclosing them; 
(2) Points on an intact rational surface;
(3) X-points such as at the tokamak separatrix or the X-points of an island chain.

The trace also determines the topological index of the fixed points, with $\Trr(\MM)<2$ having positive index and $\Trr(\MM)>2$ having negative index~\cite{greene1968two}. The trace is also connected to a quantity called Greene's residue, $\RR=\tfrac{1}{2} -\tfrac{1}{4}\Trr(\MM)$, whose sign corresponds to the topological index and whose magnitude vanishes for parabolic fixed points. 

It must be noted that the hyperbolic subgroup consists of two disjoint sets, those where $\Trr(\MM)>2$ and those where $\Trr(\MM)<-2$. 
An island chain must contain an equal number of positive-index as negative-index fixed points, so it must contain hyperbolic points of the former variety, i.e. normal X-points. 
It is however possible (and an example is shown in section \ref{sec:quasr1258083}) that the island chain does not contain O-points, but hyperbolic points with positive index. 
Such fixed points have been called hyper-hyperbolic~\cite{solov1970plasma}, alternating-hyperbolic~\cite{smiet2020bifurcations}, or M\"obiussian~\cite{wei2023invariant}.  
Confinement in such field configurations has been analyzed by Solov'ev and Shafranov~\cite{solov1970plasma}, and transition from elliptic to hyperbolic have been postulated as the cause of low $q_0$ sawtooth crashes in tokamaks~\cite{smiet2020bifurcations}.

The eigenvectors of $\MM$ describe the direction in which the transformations act e.g. in the case of X-points, the directions of stretching and squeezing. Focusing on the regular X-point ($\Trr(\MM)>2$), $\MM$ has two real and positive eigenvectors and eigenvalues $\vv_i, \lambda_i$, and because $\Det(\MM)=1$, $\lambda_1 = 1/\lambda_2$, and we choose without loss of generality $\lambda_1<1$. Thus, $\vv_1$ defines a direction in which points are mapped closer and closer to the fixed point, and $\vv_2$ the direction along which points are mapped further away (see figure \ref{fig:indexcalc}). 
The points that move away do so exponentially (provided one remains sufficiently close to the X-point that equation \eqref{eq:linearmap} holds), increasing with a factor $\lambda_2$ for each full mapping, and thus an effective Lyapunov exponent $\gamma$ \cite{wolf1986quantifying} can be calculated:
\begin{equation}\label{eq:lyapunov}
    \gamma = \log(\lambda_2).
\end{equation}
If one applies the inverse of the field line map (i.e. traces field lines in the negative toroidal direction), these behaviours are switched i.e. $\vv_1$ becomes the direction of approaching the X-point and $\vv_2$ the direction of departure.

The eigenvectors of $\MM$ also define the start of the \emph{manifolds} of a hyperbolic point. The \emph{stable manifold} is defined as the set of all trajectories whose limit is the fixed trajectory (i.e. asymptotically approaches the X-point). Near the X-point, this aligns with $\vv_1$. The stable manifold provides a boundary between 'watersheds' where trajectories are sent one side or the other, which causes the diverting character of an X-point. 
Similarly, the ~\emph{unstable manifold} is the set of trajectories whose backward limit (applying the inverse of the field line map) is the fixed trajectory. 
One can visually represent the manifolds by plotting the trajectories of a large set of points centered on the fixed point, as these all leave in the direction of the unstable manifold. Likewise, the stable manifold is obtained by calculating the backwards trajectories. We apply this methodology in sections \ref{sec:pb_nrd} and \ref{sec:quasr_db}, and an example of the methodology is shown in figure \ref{figure4}. Because magnetic field lines are toroidally continuous, X-points and their manifolds are also toroidally continuous (an X-point is sometimes called an ``X-line" for this reason). 

\subsection{Application to fields in fusion reactors}\label{sec:map_validity}
It must be noted that the domain of the Poincar\'e map in the magnetic fields of fusion devices can be limited; in general, not every field line can be followed toroidally for a whole field period. For example, field lines near magnetic coils may wind around the coils and fail to trace a field period. By definition, the region enclosed by the LGMS must be in the domain of the Poincar\'e map, but there is no general formula for how far this domain extends beyond the LGMS. For the studies presented here, it is necessary to define the \emph{maximal contractible $n$-mapping set} ($\mathcal{M}_n$), which is the largest simply-connected domain in the Poincar\'e plane in which all points can be mapped $n$ field periods. (Note that $m>n\implies \mathcal{M}_m\subseteq\mathcal{M}_n$; the set of points which complete $m$ field periods is a subset of the set of points which complete $n$ field periods for $m>n$). The domain of the Poincar\'e map is related (but not identical) to the domain within which the magnetic field is equivalent to a $1\tfrac{1}{2}$-dimensional Hamiltonian dynamical system \cite{kerst1962influence, duignan2024global}. 

Before engaging in such distinctions, however, we first examine the NRD Hamiltonian system, for which the validity of mapping and Hamiltonian theorems is assured everywhere. 

\section{Non-resonant divertor Hamiltonian topology}\label{sec:pb_nrd}
\begin{figure}
    \centering
    \includegraphics[width=1.0\linewidth]{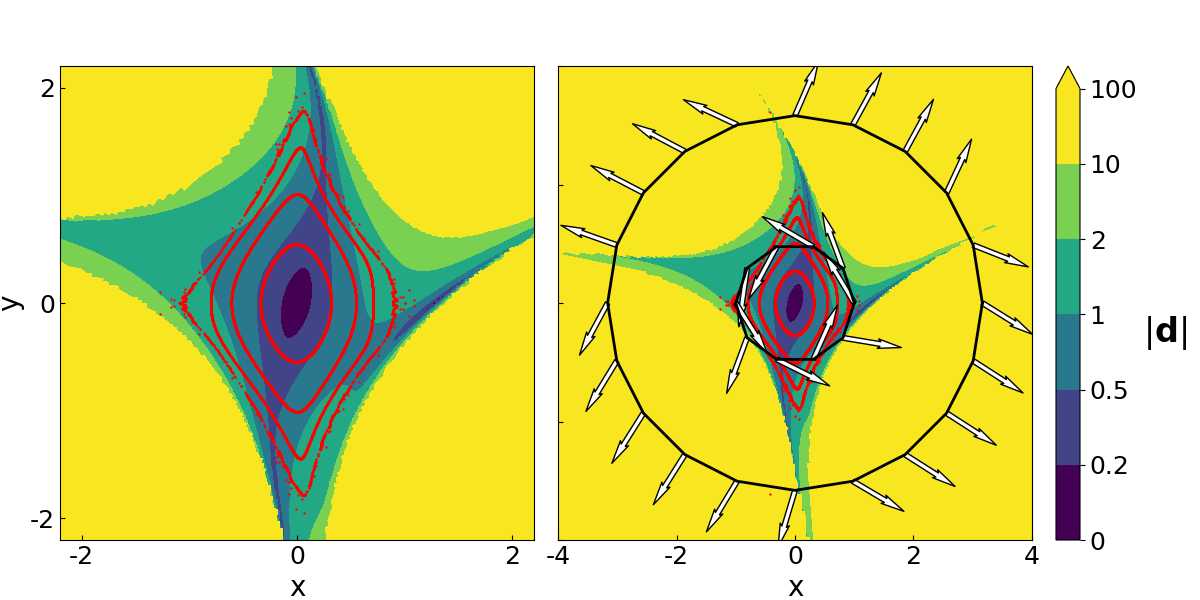}
    \caption{Poincar\'e section for the non-resonant divertor Hamiltonian and displacement vector magnitude for the forwards map, $\vert \dd_f \vert $ at toroidal location $\xi=0$. Left: a zoom-in of the confined region. Right: arrows indicate direction of $\dd_f$ at locations on closed contours (black lines), over which the winding number is calculated.}
    \label{figure1}
\end{figure}

\begin{figure}
    \centering
    \includegraphics[width=1.0\linewidth]{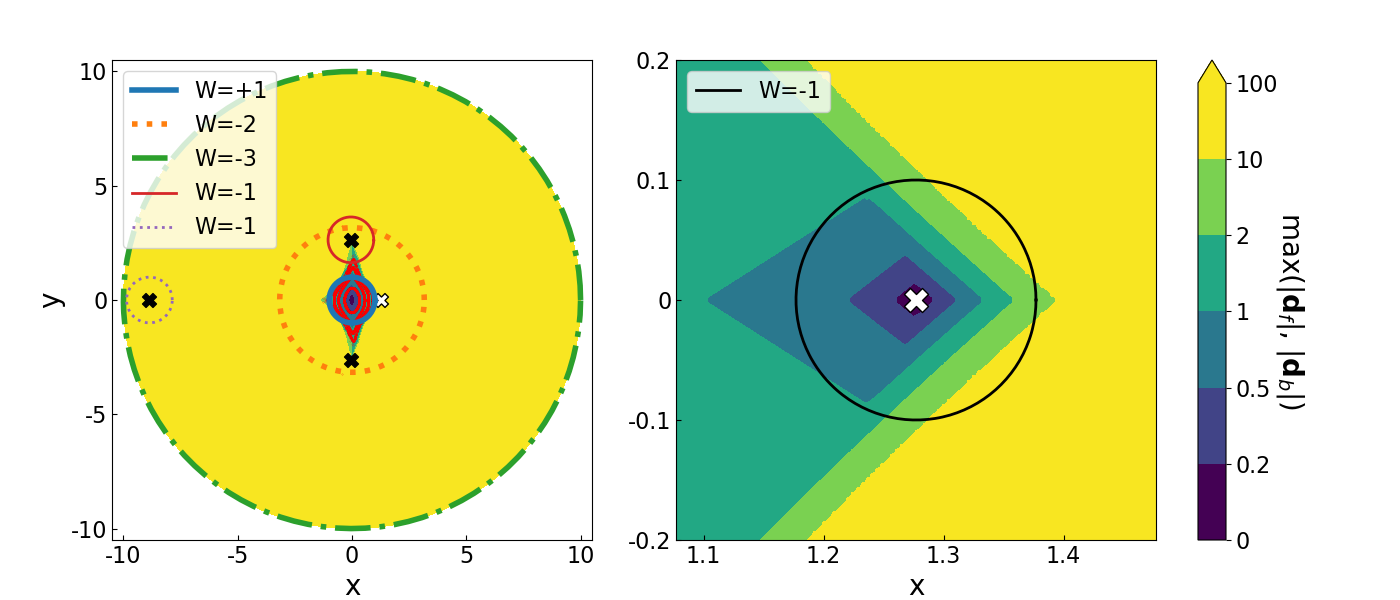}
    \caption{Maximum displacement of the forwards and backwards maps and closed contours over which winding numbers are calculated for the non-resonant divertor Hamiltonian. The left plot shows all four X-points of the system. The right plot shows a zoom-in of the innermost X-point (i.e. the X-point with the lowest $\psi_t$), depicted as a white cross.}
    \label{figure3}
\end{figure}
\begin{figure}
    \centering
    \includegraphics[width=1.0\linewidth]{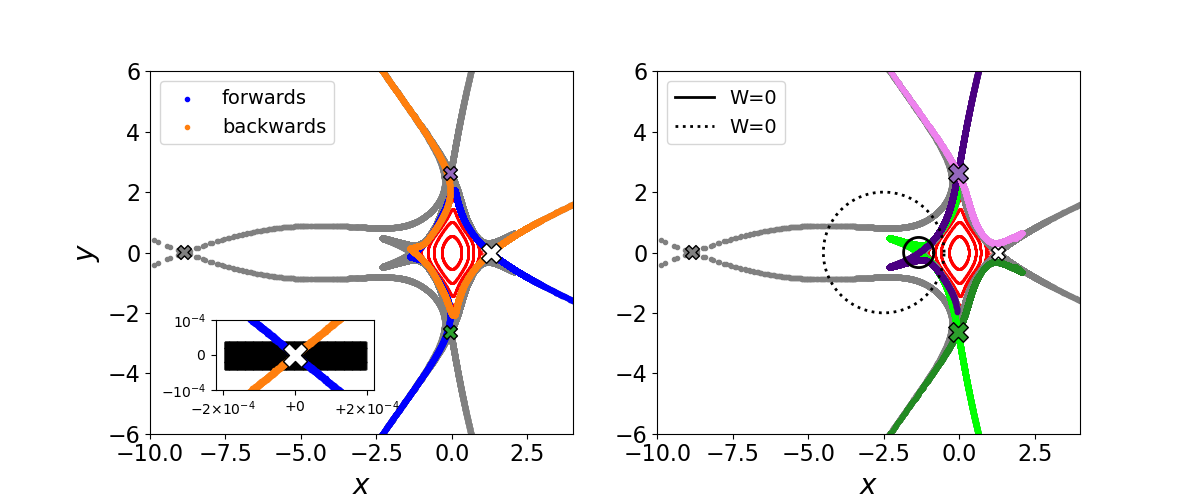}
    \caption{Manifolds of the four X-points of the NRD Hamiltonian system shown in grey, blue, orange, green and purple. Left: manifolds of the innermost X-point when tracing forwards in toroidal position $\xi$ (blue) and backwards in $\xi$ (orange). These are found by tracing a block of $300 \times 30$ points which are initialised around the X-point. This dense blob of points is shown in black in the inset plot, lower left, The x-axis ticks are given with respect to the X-point location. Right: Manifolds of the upper (light purple and dark purple) and lower (light and dark green) X-points, with light/dark representing forwards/backwards tracing, calculated in a similar way. Also includes closed contours used for winding number calculations over the region where these manifolds make X-shaped crossings. These show that this feature is not an X-point.}
    \label{figure4}
\end{figure}
\begin{figure}
    \centering
    \includegraphics[width=1.0\linewidth]{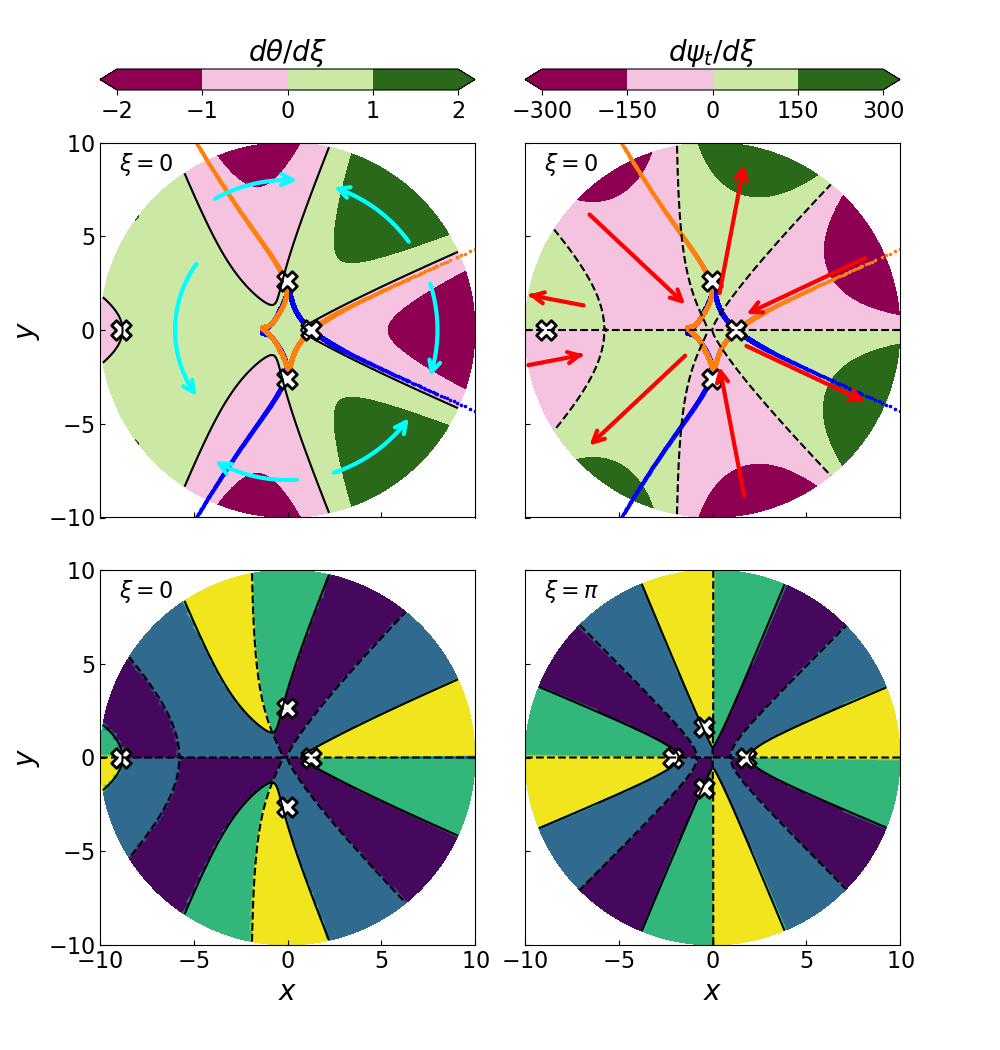}
    \caption{Derivative information for the NRD Hamiltonian. Upper row: poloidal velocity $\dthetadxi$ at $\xi=0$ (left) and radial velocity $\dpsidxi$ at $\xi=0$ (right), also showing arrows indicating the direction of poloidal and radial travel as one advances in $\xi$, and manifolds of the innermost X-point. Bottom row: Color-coded regions for the different signs of $\dthetadxi$ and  $\dpsidxi$ at toroidal locations $\xi=0$ (left) and at $\xi=\pi$ (right).}
    \label{figure5}
\end{figure}

The non-resonant divertor Hamiltonian system \cite{boozer2018simulation, punjabi2020simulation, punjabi2022magnetic} consists of a Hamiltonian of the form:
\begin{align}
    \frac{\psi_p}{\bar{\psi}_g} = &\left [\iota_0 + \frac{\epsilon_0}{4} ((2\iota_0-1)\cos(2\theta-\xi) +2\iota_0\cos(2\theta)) \right] \left (\frac{\psi_t}{\bar{\psi}_g} \right) \nonumber \\
    &+ \frac{\epsilon_x}{8} \left [(4\iota_0-1)\cos(4\theta-\xi) +4\iota_0\cos(4\theta) \right] \left (\frac{\psi_t}{\bar{\psi}_g} \right)^2 \nonumber \\
    &+ \frac{\epsilon_t}{6} \left [(3\iota_0-1)\cos(3\theta-\xi) +3\iota_0\cos(3\theta) \right] \left (\frac{\psi_t}{\bar{\psi}_g} \right)^{3/2}, \label{eq:PBNRD_hamiltonian}
\end{align}
with trajectories given by
\begin{align}
    \dpsidxi  &= - \frac{\partial \psi_p}{\partial \theta}, \label{eq:dpsitdphi}\\
    \dthetadxi &= \frac{\partial \psi_p}{\partial \psi_t}. \label{eq:dthetadphi}
\end{align}
Here, $\psi_p$ is the Hamiltonian of the system and ($\psi_t, \theta, \xi$) are the phase-space coordinates (representing toroidal flux, poloidal angle and toroidal angle respectively). $\xi$ is the time-like coordinate, acting as the toroidal angle within a single field period of the stellarator, and is related to the ordinary toroidal angle $\varphi$ by the number of field periods: $\xi = n_\text{fp}\varphi$. Each field period is taken to be strictly identical, so the dynamics of the system can be understood by examining a single field period, which extends from $\xi=0$ to $\xi=2\pi$. $\bar{\psi}_g$ is a normalising quantity representing a characteristic toroidal magnetic flux. 

The quantities $\iota_0$, $\epsilon_0$, $\epsilon_t$ and $\epsilon_x$ are input parameters used to adjust the properties of the system; $\iota_0$ is the rotational transform of the magnetic axis over a single field period and $\epsilon_{0,t,x}$ affect the shape of flux surfaces. In this work we select the same parameters as Punjabi and Boozer \cite{punjabi2020simulation, punjabi2022magnetic}
: $\iota_0=0.15$, $\epsilon_0 = \epsilon_t = 0.5$, $\epsilon_x=-0.31$, $n_\text{fp}=5$. It should be noted that since $\varphi$ does not explicitly appear in equations (3-5), the dynamics of the system are unaffected by the choice of $n_\text{fp}$.

The coordinates $(\psi_t, \theta)$ are taken to be related to real-space coordinates ($x,y$) by 
\begin{align}
    x &= \sqrt{\psi_t/\bar{\psi}_g} \cos(\theta), \\
    y &= \sqrt{\psi_t/\bar{\psi}_g} \sin(\theta).
\end{align}
We implement the system of equations (3-5) in python using the Scipy odeint library. 

A Poincar\'e section at $\xi=0$ is shown in figure \ref{figure1}, left, with initial points selected by scanning $\psi_t$ until good flux surfaces cease to exist. Trajectories started a small distance beyond the LGMS stay near the LGMS for several field periods. Trajectories initialised slightly further away are quickly diverted  away from the LGMS, often before travelling a single field period in $\xi$. This behaviour has been attributed to a cantorus near the LGMS \cite{boozer2018simulation, punjabi2020simulation, punjabi2022magnetic}, that is, the remnant of an irrational surface which acts as a transport barrier. If this were a stellarator and a wall were placed around the confined region, heat and particles expelled from the LGMS would travel along these outgoing trajectories until they intersect the wall.

We define the ``forwards map", $f_f(\xx)$, the result integrating equations (3-5) by $2\pi$ in $\xi$, and a ``backwards map"  $f_b(\xx)$ as an integration of $-2\pi$ in $\xi$. We define the displacement vector $\dd_{f}(\xx) = f_f(\xx) - \xx$, where $\xx = (x,y)$ is the initial position vector in the toroidal plane and $f_f(\xx)$ is the position vector after applying the forwards map. 
Similarly, $\dd_{b}(\xx)$ is defined as the displacement vector associated with the backwards map. $\dd_{f}=0$ indicates a fixed point of the map (for which, necessarily, $\dd_{b}=0$). The magnitude of $\dd_f$ and arrows indicating direction is shown in figure \ref{figure1}, right. It can be seen that $\vert \dd_f \vert$ quickly becomes large beyond the LGMS, consistent with the trajectories being quickly diverted away from the confined region. The direction of the vector on the two black contours show qualitatively different behaviour. The inner ring (located at $\psi_t/\bar{\psi}_g=1$) shows the vector rotating fairly uniformly as one follows the contour, whereas in the outer contour ($\psi_t/\bar{\psi}_g=10$) shows sharp changes at certain locations. 

We use the forwards map to calculate the winding number $W$ over selected closed loops at $\xi=0$, $\psi_t/\bar{\psi}_g=(1,10,100,1000)$, the first two of which are shown in figure \ref{figure1}, right and the first three in figure \ref{figure3}, left. Because the direction of $\dd_f$ varies rapidly at certain locations, we adjust the step size $d\theta$ of successive points to ensure the angle of $\dd_f$ changes by less than $2\pi/100$ between successive points. The winding numbers for these four loops are $W=(+1, -2, -3, -3)$ respectively. This calculation proves the existence of fixed points with negative topological index between the first and second closed contours ($\psi_t/\bar{\psi}_g=1$ and $\psi_t/\bar{\psi}_g=10$) and between the second and third contours ($\psi_t/\bar{\psi}_g=10$ and $\psi_t/\bar{\psi}_g=100$). In total, there must be three more negative-index fixed points (X-points) than positive-index fixed points (e.g. O-points) within the region bounded by $\psi_t/\bar{\psi}_g=100$.

These winding numbers only provide information about the single-field period map. Repeating the analysis, applying the map for two field periods, yields the same winding numbers; we find $W=(+1, -2, -3, -3)$ for $\psi_t/\bar{\psi}_g=(1,10,100,1000)$. From this we conclude that either there are no additional fixed points with this periodicity, or that there are additional O- and X-points, but they exist in equal number in between each of our contours. However, the particular properties of this Hamiltonian (discussed presently) lead us to expect that, beyond a particular $\psi_t/\bar{\psi}_g$, fixed points can only have a rotational transform of zero.

We find the single-field period fixed points of the system manually, and then verify them using winding number calculations over small loops containing the fixed points. Our search is performed by plotting the maximum of the magnitude of the forward and backward displacement vectors at toroidal location $\xi=0$ and finding locations where this approaches zero. This quantity is plotted in ($x,y$) space in figure \ref{figure3}, left, which also shows the locations of the X-points, and contours around two of the X-points, for each of which we find $W=-1$. By following tracing the X-points over one field period (from $\xi=0$ to $\xi=2\pi$) we find that the X-points have a rotational transform of zero. 

A zoom-in on the innermost X-point (that is, the X-point with the smallest radial coordinate $\psi_t$) is shown in \ref{figure3}, right. Because this is the X-point nearest to the LGMS, it is reasonable to expect that this is the most important for diverting trajectories started near to the LGMS. We trace out the manifolds of this X-point by mapping a dense blob of points for several field periods in each direction. The results are shown in figure \ref{figure4}, left, with forwards and backwards trajectories shown in blue and orange respectively, and the manifolds of the other X-points (calculated in the same way) shown in grey. In each direction, there are two distinct bundles of diverted trajectories, consistent with the earlier results \cite{punjabi2022magnetic}. 

It is worth noting that there is a region of chaos in between the LGMS and the X-point; that is, good magnetic flux surfaces do not extend all the way to the innermost X-point. The transport induced by chaos in regions where good surfaces do not exist \cite{meiss1992symplectic}, as well as cross-field transport is likely to affect the relative importance of the X-points. We do not study this here but note that it would be a valuable future study. 'Diversion', i.e. the complete theory of transport from plasma edge to the wall, and the relative amount of plasma carried to the wall from each of the X-points depends crucially on these chaotic interactions and has been partially studied in the Hamiltonian system [9] but is an interesting area for future research.

Figure \ref{figure4}, right, shows the manifolds of the upper and lower X-points, which intersect each other to form an X-shaped pattern around ($x=-2, y=0$) (previously observed and labeled ``pseudo-turnstiles" \cite{punjabi2022magnetic}). We note that although the pattern is X-shaped, it is \textit{\textbf{not}} an X-point (it is not a fixed point of the Poincar\'e map), but rather an intersection of manifolds known as a heteroclinic intersection \cite{meiss1992symplectic}. 

To understand why the four X-points in this system come about, we compute the terms $\dthetadxi$ and $\dpsidxi$, which can be thought of as the poloidal and radial ``velocities" of the trajectories. This is shown in figure \ref{figure5}. The upper left plot reveals surfaces on which $\dthetadxi = 0$, which correlate with the X-points and their manifolds. These surfaces arise because the Hamiltonian \eqref{eq:PBNRD_hamiltonian} contains $\theta$-dependent terms which grows faster-than-linearly with $\psi_t$ and thus the poloidal velocity $\dthetadxi$ changes sign at certain values of $\theta$ for sufficiently large $\psi_t/\bar{\psi}_g$. 

The change in sign of poloidal velocity means that trajectories beyond a critical $\psi_t/\bar{\psi}_g$ cannot complete a full poloidal rotation. Instead, the forwards-traveling trajectories are attracted to the contours on which $\dthetadxi=0$ and $\partial / \partial \theta (\dthetadxi) < 0$ (i.e. where the blue arrows in figure \ref{figure5}, upper left, point at one another) and the backwards-traveling trajectories are attracted to contours where $\dthetadxi=0$ and $\partial / \partial \theta (\dthetadxi) > 0$. The radial motion $\dpsidxi$ (figure \ref{figure5}, upper right) pushes the trajectories either radially inwards or outwards, and fixed points appear where both the poloidal and radial motions vanish, $\dthetadxi=\dpsidxi=0$. This picture is slightly complicated by the fact that the contours of $\dthetadxi=0$ and $\dpsidxi=0$ are $\xi$-dependent, causing the X-points to move as $\xi$ changes. To show this, the contours and the X-points at two toroidal locations, $\xi=0$ and $\xi=\pi$, are shown in figure \ref{figure5} (lower row). It is the change in sign of poloidal velocity which effectively forbids fixed points with $\iota \neq 0 $ beyond a certain $\psi_t/\bar{\psi}_g$.

These findings show that diversion is largely determined by the manifolds of ``unpaired" X-points; analogously to tokamaks, X-points exist just beyond the LGMS, and trajectories are quickly deviated by their manifolds. If these X-points were part of an island chain, their manifolds would encircle O-points and the diverting trajectories would only have a finite deviation from the confined region. However, the NRD Hamiltonian is such that corresponding O-points are not present in the system. (This can be seen by the winding number calculations, and by observing that the sign of radial velocity $\dpsidxi$ for the manifolds cannot vary when $\psi_t$ is sufficiently large.) Because the X-points are not paired with O-points, their manifolds extend arbitrarily far from the confined region.

\begin{figure}
    \centering
    \includegraphics[width=0.8\linewidth]{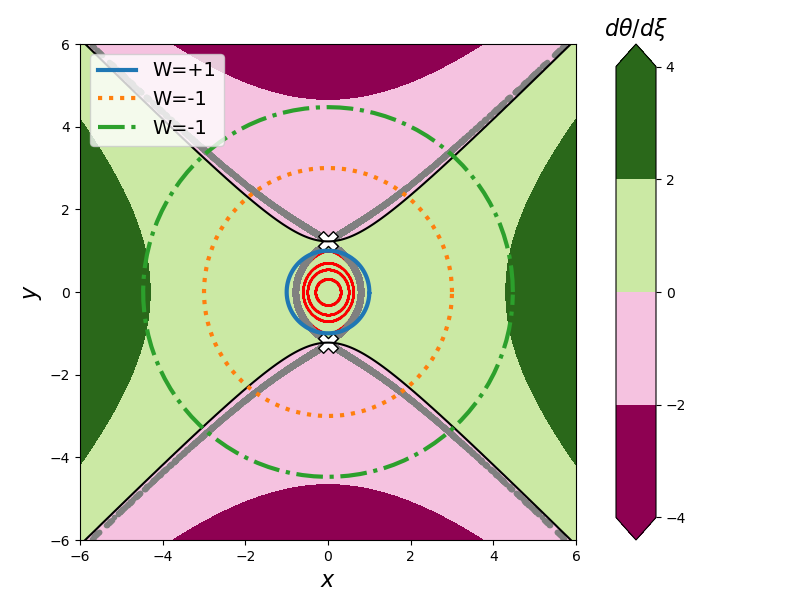}
    \caption{Properties of the axisymmetric Hamiltonian \eqref{eq:axisymmetric_hamiltonian} system with similar properties to the NRD Hamiltonian system. Poloidal velocity $\dthetadxi$ is shown in filled colours and the contours over which the winding number is calculated are shown by blue, orange and green lines. A Poincar\'e section is shown in red, the X-points by white crosses and X-point manifolds shown in grey.}
    \label{figure7}
\end{figure}

To examine whether this phenomena explicitly relies on 3D effects, we show the same behaviour in a similar, axisymmetric Hamiltonian system:
\begin{align}
 \frac{\psi_p}{\bar{\psi}_g} &= \iota_0  \frac{\psi_t}{\bar{\psi}_g} + A\left( \frac{\psi_t}{\bar{\psi}_g}\right)^2\cos(2\theta), \label{eq:axisymmetric_hamiltonian}
\end{align}
with $\iota_0 = 0.15$, $A=0.05$. As in eq. \eqref{eq:PBNRD_hamiltonian}, there is a $\theta$-dependent term which grows faster-than-linearly with $\psi_t$. Thus, there is a $\theta$-dependent perturbation to the poloidal velocity which grows with $\psi_t$, and creates planes on which the poloidal velocity vanishes above a critical $\psi_t$.

The topological information of this system is summarised in figure \ref{figure7}. We find the winding numbers on surfaces with constant $\psi_t/\bar{\psi}_g=(1,9, 20)$ are ($+1$, $-1$ $-1$) respectively, indicating two X-points in the region ($1 < \psi_t/\bar{\psi}_g < 9$). These are located and shown to coincide with $\dthetadxi=\dpsidxi=0$. This is expected; trajectories at $\dthetadxi=\dpsidxi=0$ will have no poloidal or radial motion, and since $\dthetadxi$ and $\dpsidxi$ do not vary with $\xi$, these will be fixed points. As in the NRD case, the X-point manifolds lie close to the $\dthetadxi=0$ contour, showing that the diversion could be predicted by examining the derivatives of the Hamiltonian.

This study illustrates the importance of topology on diversion, and specifically how unpaired X-points (X-points which are not part of an island chain) can cause arbitrarily large diversion of Hamiltonian trajectories. This prompts the question: can unpaired X-points be present in a buildable stellarator? LHD could be seen one example of this phenomena - X-points are generated by helically rotating coils. One could interpret these coils as O-points which ``pair" with the X-points, but it is unclear whether the domain of the Poincar\'e map (and therefore the domain in which one can meaningfully discuss fixed points) extends this far.

Apart from the special geometry of LHD, are unpaired X-points a general feature of realistic stellarators? We address this in the next study, by searching the QUASR database for such configurations. 

\section{Topological search of the QUASR database}\label{sec:quasr_db}
QUASR (`quasi-symmetric stellarator repository') \cite{giuliani2024direct, giuliani2024comprehensive} is a repository of over $300,000$ quasi-axisymmetric and quasi-helically symmetric stellarator vacuum configurations, including the coil sets which produce them. Quasisymmetry is one of several methods of minimising neoclassical transport, which is necessary for stellarator reactors \cite{helander2014theory}. 

The magnetic field beyond the plasma boundary was not explicitly considered, constrained or optimised when the database was created, and therefore this work could be considered a ``fishing expedition" to see what edge topologies can naturally arise in stellarators. We do not model any plasma dynamics in these configurations apart from their vacuum magnetic characteristics. Establishing whether they meet the requirements of an exhaust solution for a commercial power plant would require many more steps (PFC design, transport modelling, etc). Such studies are beyond the scope of this investigation, which simply examines the topology of a fixed magnetic field.

Our search is performed as follows. For a given configuration, the coils are retrieved from the database, and the Biot-Savart law is used to calculate the magnetic field on a regular ($R, \phi, Z$) grid. The $R$ and $Z$ extent of the grid are set to be the most extreme $(R,Z)$ of the coil locations - that is, the grid is the smallest box which contains all of the coils, whilst having a constant rectangular cross-section at each toroidal location $\phi$. The number of field periods for the map, $n_\text{map}$, are decided by the user - for example, $n_\text{map}=4$ will give information of all the fixed points of the four field period map (which also includes the fixed points of its divisors i.e. fixed points of the one field period map and the two field period map). Magnetic field lines are traced for $n_\text{map}$ field periods in order to calculate the winding number over the maximal contractible $n_\text{map}$-mapping set (see section \ref{sec:map_validity} for definition). Finding the maximal contractible $n_\text{map}$-mapping set and ensuring sufficient resolution of points is automated in the following way.

We first trace a regular grid of $(R,Z)$ points with resolution $dR=dZ=1\text{cm}$ over the entire domain on which the magnetic field is calculated. The field line tracer integrates along a field line, aborting under two conditions: (1) the trajectories leave the $(R,Z)$ domain on which the magnetic field is calculated; or (2) the toroidal magnetic field changes sign (i.e. the magnetic field line changes toroidal direction, as can happen in the neighbourhood of the coils). We record the trajectories which traverse $n_\text{map}$ field periods and find the largest closed contour of these points. For this contour, the change in direction of $\dd_f$ between successive points on the contour is calculated. The ($R,Z$) points on the contour is upsampled by linear interpolation wherever the change in direction of $\dd_f$ exceeds a tolerance of $2\pi/100$. The largest contour is recalculated each time new points are added to ensure that the contour remains well-behaved. The upsampling is repeated until the phase change is below the tolerance for all successive points on the largest contour. The total computational cost per configuration (downloading the data, running the Biot-Savart calculation and running the winding number calculation) is around 100 seconds on a single CPU.

We perform two sets of searches. Firstly, we apply our method to the first $20$  single-field period stellarators in the database with $n_\text{map}=1$. Secondly, we scan approximately $100$ four-field period stellarators with an average rotational transform in the range $0.8 \leq \iota \leq 1.2$, using with $n_\text{map}=4$. From these two searches, $34$ reported $W<1$ and two reported $W>1$, indicating that the database is rich with potentially interesting topologies. 
We select three configurations for further analysis. For these cases, we also use the pyoculus package \cite{pyoculus} to calculate the Jacobian $\MM$ of the fixed points, with the results reported in \ref{app:pyoculus}.

\subsection{Tokamak-like diversion: configuration 104183}\label{sec:quasr104183}
\begin{figure}
    \centering
    \includegraphics[width=1.0\linewidth]{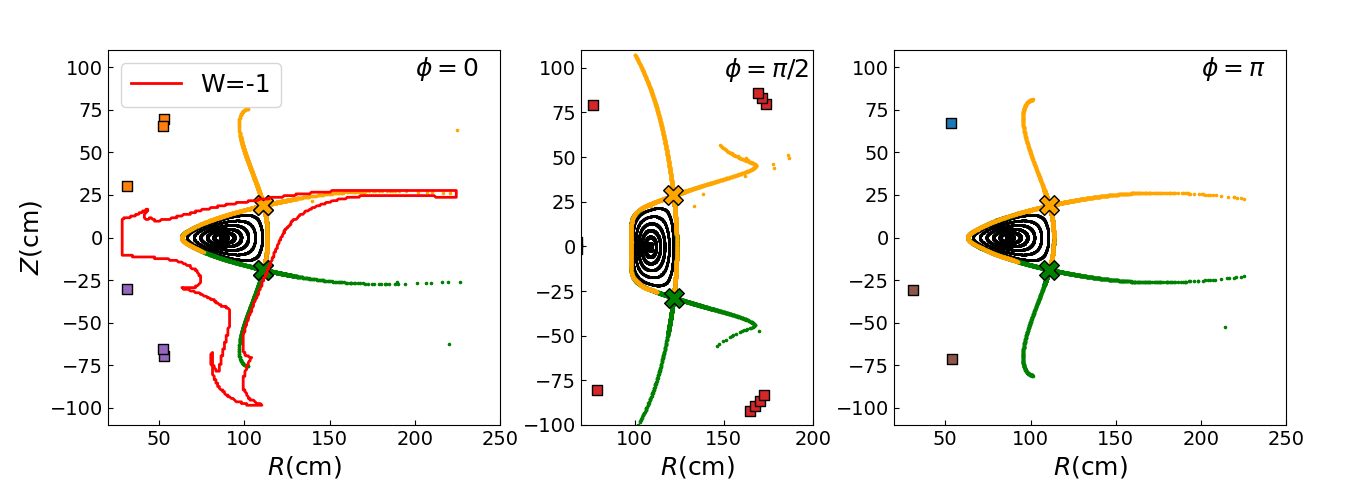}  
    \caption{X-points and manifolds for (quasi-axisymmetric) QUASR configuration 104183 at three toroidal locations ($\phi=0, \pi/2, \pi$). Poincar\'e section is shown in black and the coil locations shown as colored squares (each square represents a point on the coil which is within $2.5^\circ$ of this toroidal location). Leftmost plot shows the closed contour over which the winding number $W$ is calculated (red line). $W=-1$ over the contour proves that there are two more X-points than O-points within the contour. These X-points are found and plotted as green and orange crosses and their manifolds as green/orange dots. }
    \label{figure_quasr104183}
\end{figure}
Configuration 104183 is is a single-field period quasi-axisymmetric device with a low value of rotational transform ($\iota=0.21$ on the magnetic axis and $\iota=0.19$ near the LGMS). Such a low value might not be desirable in a reactor. However, this serves as a first example of unpaired X-points in the QUASR database.

We calculate $W=-1$ over the the maximal contractible $1$-mapping set, and manually find two X-points, which are shown in figure \ref{figure_quasr104183}. The X-points have $\iota=0$, and always remain at the top and bottom on the outboard side. It should be noted that although the stellarator is single-field period, each half-field period is nearly identical so looks rather like a two-field period device (as shown in figure \ref{figure_quasr104183}). Such behaviour is not unexpected \cite{giuliani2024direct}.

The maximal contractible $1$-mapping set is also shown in figure \ref{figure_quasr104183}. The contour stretches over the stable manifolds of the X-points (i.e. the manifolds which approach the X-point when tracing in the positive $\phi$ direction) and compresses over the unstable manifolds. A consequence of this is that the X-points lie close to the contour. This makes the spatial resolution of the initial grid important, and is the reason we select the relatively high-resolution parameters $dR=dZ=1\text{cm}$. 

The manifolds show that these X-points guide the magnetic field away from the confined region. This is qualitatively similar to the non-resonant divertor Hamiltonian or a double-null tokamak configuration. The latter is surprising, since a plasma current is necessary for tokamak X-points but is not required here.

\subsection{Unpaired X-points with $\iota=1$: configuration 74609}
\begin{figure}
    \centering
    \includegraphics[width=1.0\linewidth]{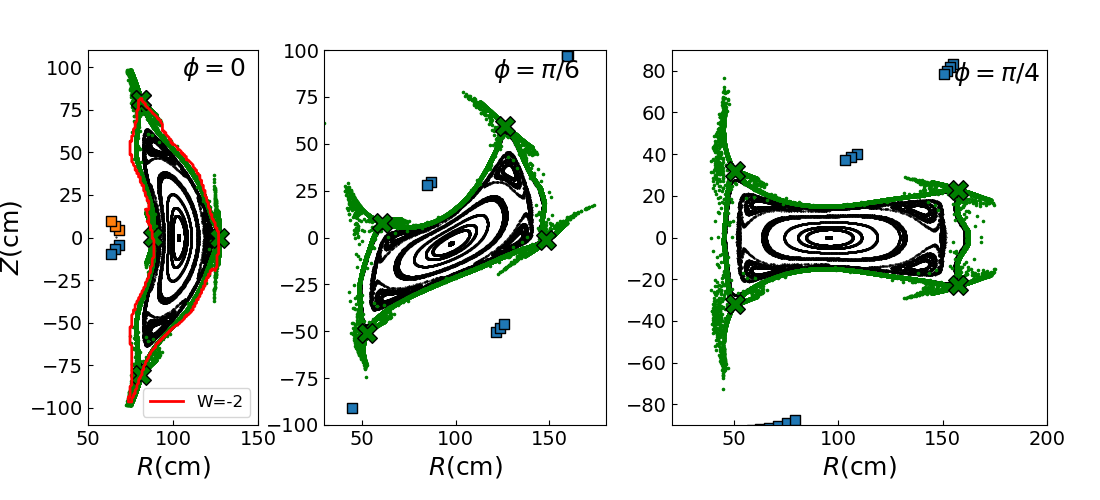}
    \caption{Poincare (black points) and contour of the maximal mapping set (red line) for (quasi-axisymmetric) QUASR configuration 74609 at three toroidal locations ($\phi=0, \pi/6, \pi/4$). This configuration shows four unpaired X-points with rotational transform $\iota=1$, displayed as green crosses. Manifolds are shown as green dots.}
    \label{figure_quasr74609}
\end{figure}
Can unpaired X-points exist in configurations with higher values of rotational transform? This is shown by configuration 74609, illustrated by figure \ref{figure_quasr74609}. This stellarator is a quasi-axisymmetric stellarator with four field periods, relatively low aspect ratio ($A=5.0$) and a rotational transform varying from $\iota=0.86$ (magnetic axis) to $\iota=0.95$ near the QUASR-identified LGMS. We find $W=(-2)$ for $n_\text{map}=4$. Further investigation shows that this is caused by a $4/4$ chain of X-points which do not have ``compensating"  O-points within the maximal contractible $4$-mapping set. The winding number is $W=-2$ rather than $W=-3$ because the spatial resolution of the contour causes it to miss one of the X-points. 

The manifolds divert trajectories from the confined region, albeit not as cleanly as configuration 104183. It is also interesting to note that, in addition to these outermost $4/4$ X-points, an ordinary $4/4$ island chain is present (implying that the rotational transform profile in the edge is non-monotonic). An interesting piece of future work would be to examine the variation of both $4/4$ chains of fixed point with equilibrium effects; are the unpaired X-points more robust than those of the inner island chain?

\subsection{Interacting $\iota=0$ and $\iota=1$ fixed points: configuration 1258083}\label{sec:quasr1258083}
\begin{figure}
    \centering
    \includegraphics[width=1.0\linewidth]{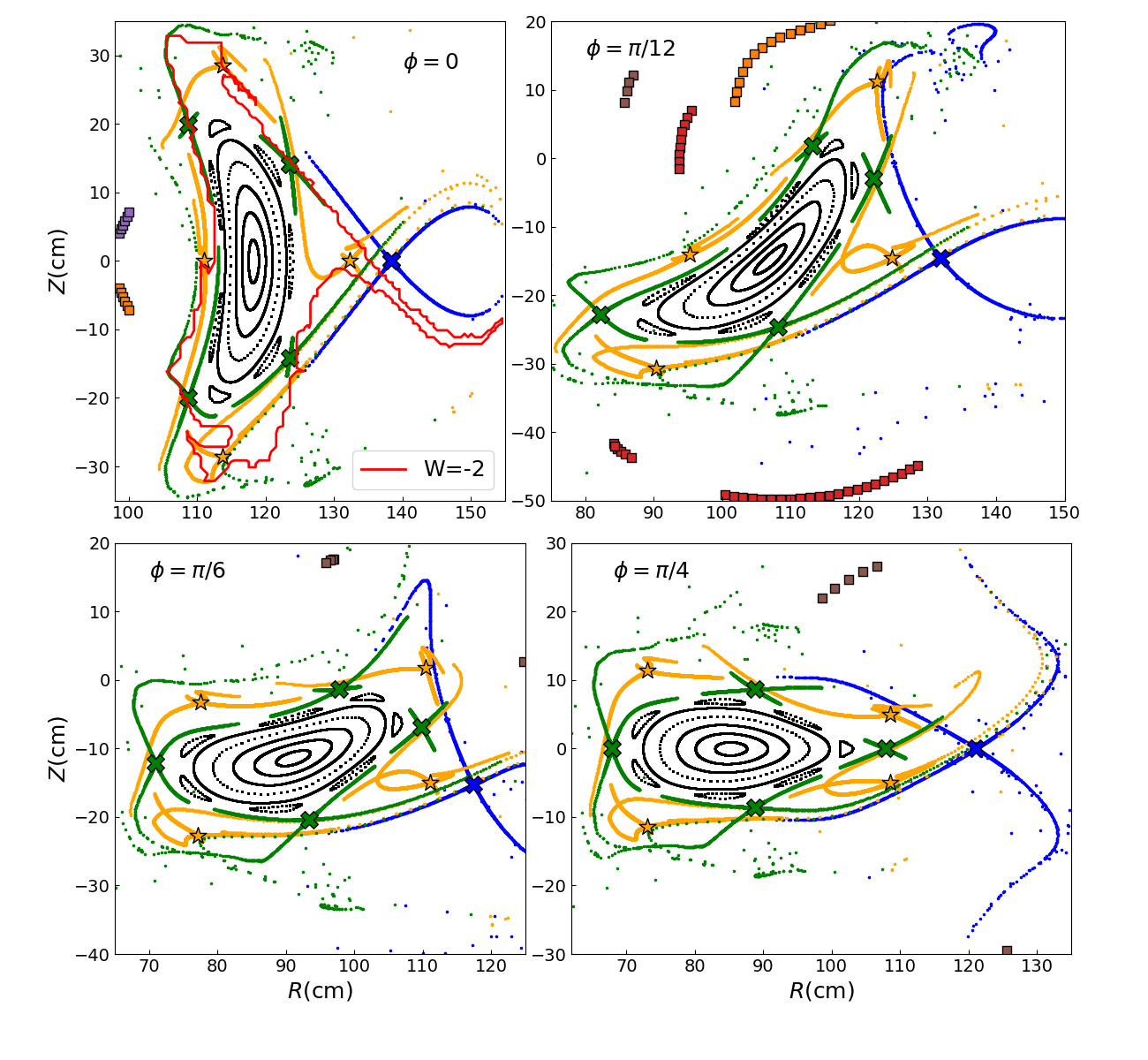}
    \caption{Poincare plot (black) and fixed point information for the (quasi-helically symmetric) QUASR configuration 1258083 at four toroidal locations. This configuration contains a 4/4 island chain consisting of regular X-points (green crosses) and positive-index hyperbolic points (orange stars). The configuration also contains an X-point with rotational transform $\iota=0$ (blue cross). Manifolds are shown as colored dots.}
    \label{figure_quasr1258083}
\end{figure}
Finally, we show analysis for configuration 1258083, a four-field period quasi-helically symmetric stellarator, in figure \ref{figure_quasr1258083}. This configuration is probably unattractive as a fusion reactor due to its high aspect ratio ($A=24.0$) and strongly shaped magnetic coils. However, this configuration displays interesting topological features, illustrating the breadth of novelty available to stellarators. 

We find $W=-2$ for $n_\text{map}=4$. Further analysis identifies three types of fixed points in the edge: (1) a chain of $\iota=1$ X-points (green in figure \ref{figure_quasr1258083}),  (2) a chain of $\iota=1$, positive-index hyperbolic fixed points, known as ``hyper-hyperbolic fixed points" (see section \ref{sec:jacobian}) (orange in figure \ref{figure_quasr1258083}) and (3) an $\iota=0$ X-point located slightly further from the confined region (blue in figure \ref{figure_quasr1258083}). In practice, the relative importance of these fixed points, like in the NRD Hamiltonian example, would depend on the transport established by the chaotic region, as well as the cross-field plasma transport and the location of the PFCs. The $\iota=0$ fixed point appears to reliably divert field lines to the outboard side of the vessel. If such a feature is robust to equilibrium changes, and can be engineered without sacrificing too much confined plasma volume (as appears to be the case here), it might provide an acceptable exhaust solution, with PFCs located on the outboard side of the machine.

\section{Conclusions}\label{sec:conclusions}
This work uses winding number calculations of the field line map to analyse the number and nature of fixed points in stellarator edges. Specifically, we study ``unpaired" X-points (X-points which are not part of an island chain and therefore not ``paired" with an equal number of O-points), and how these guide field lines away from the confined region. We first show that unpaired X-points are present in the non-resonant divertor (NRD) Hamiltonian system. We then show examples of stellarator configurations which contain unpaired X-points. 

In the NRD Hamiltonian study, we find four unpaired $\iota=0$ X-points, the manifolds of which guide trajectories away from the confined region. The X-points occur approximately where $\dthetadxi=\dpsidxi=0$ and the manifolds correlate with surfaces on which $\dthetadxi=0$. Thus, examining the derivatives of the Hamiltonian, $\partial \psi_p/\partial \psi_t = \dthetadxi $ and $\partial \psi_p/\partial \theta = -\dpsidxi$ can offer a physical understanding and rough prediction of the system's diversion. These findings offer a new perspective on diversion in the NRD Hamiltonian system, in which unpaired X-points play a central role. 

The manifolds of these X-points define the topological structure of the field, i.e. where magnetic field (and hence to leading order the plasma) departs from the confined volume. We again emphasise that this study ignores (for example) cross-field transport, neutral and radiation physics, and the location of the plasma-facing components (PFCs). The latter is required to understand the actual locations of plasma-surface interaction; the reactor wall must be designed to intersect these structures and geometrically optimised such that material limits are not exceeded. The topological understanding of diversion will allow us to better design the wall and plasma facing components. In experimental and theoretical studies, 'helical stripes' are observed along the wall (e.g. \cite{strumberger1992magnetic, bader2017hsx, garcia2024resilient}), the exact shape of which is still an open area of study. However, the topology gives a simple metric by which to classify and study stellarator magnetic fields. 
 
In our second study, we show that unpaired X-points are a general feature of realistic stellarators. We do this by searching the QUASR database using an automated winding number calculation scheme. This approach is fast and imperfect; for example, the contour over which we calculate the winding number (i.e. the maximal contractible $n$-mapping set $\mathcal{M}_n$, see section \ref{sec:map_validity}) depends upon spatial resolution and must be interpreted carefully. We also stress that we have analysed only a tiny fraction of stellarators in the database ($\approx 100$ out of $\approx 300,000$). Nevertheless, we find several examples of unpaired X-points. Further automation and analysis of more configurations is planned future work, and may eventually provide a ``recipe" for how to engineer non-resonant divertors and perform topological optimisation.

We analyse three examples in greater detail. The first case resembles a tokamak double-null divertor, containing $\iota=0$ X-points at the top and bottom of the device without compensating O-points. However, unlike a tokamak, these X-points do not require a plasma current. The second example is a low-aspect ratio quasi-asymmetric stellarator with a reasonable $\iota$ profile and with four unpaired $\iota=1$ X-points. The resiliency of these with respect to e.g. plasma current would be a valuable area of future research. The final case shows an unpaired $\iota=0$ X-point on the outboard side of the device, interacting with a $4/4$ island chain. Additionally, both the positive- and negative-index fixed points of the $4/4$ island chain  have a hyperbolic geometry - that is, the O-points are replaced by hyperbolic fixed points. The consequences of this for heat and particle transport are unclear. These phenomena are highly novel and demonstrate that the topological freedom in buildable stellarators is greater than has currently been explored.

These findings prompt several questions on both non-resonant divertors and on the behaviour of finite-domain Poincar\'e maps: (1) Are all ``NRD" examples characterised by unpaired X-points, or only a subset? (2) Are unpaired X-points inherently more (or less) resilient to changes in the magnetic field than the fixed points of island chains? (3) If the winding number of $\mathcal{M}_n$ is less than +1 (there are the same number of/more X-points as/than O-points), does that imply that the manifolds of these unpaired X-points leave $\mathcal{M}_n$, providing efficient diversion? (4) What can be said about fixed points outside of $\mathcal{M}_n$? One might wish to interpret unpaired fixed as part of an island chain, but the associated O-points do not need to exist, as exemplified in the Hamiltonians of section~\ref{sec:pb_nrd}. 

In conclusion, we show the benefits of topological studies of stellarator diversion. Further studies and incorporating topological methods into optimisation schemes could ultimately improve the commercial viability of stellarator power plants.

\section{Acknowledgements}
This research was supported in part by the Helmholtz International Laboratory for Optimized Advanced Divertors in Stellarators (HILOADS), and by a grant from the Simons Foundation\ (1013657, JL), and by the U.S. Department of Energy, Office of Science under Award Nos. DE-FG02-95ER54333, DE-SC0024548, and DE-AC02-09CH11466.

This work has been carried out within the framework of the EUROfusion Consortium, partially funded by the European Union via the Euratom Research and Training Programme (Grant Agreement No 101052200 — EUROfusion). The Swiss contribution to this work has been funded by the Swiss State Secretariat for Education, Research and Innovation (SERI). Views and opinions expressed are however those of the author(s) only and do not necessarily reflect those of the European Union, the European Commission or SERI. Neither the European Union nor the European Commission nor SERI can be held responsible for them.
\section*{References}
\bibliographystyle{vancouver}
\bibliography{references}

\begin{thebibliography}{10}

\bibitem{soboleva1997energy}
Soboleva T.
\newblock Energy Exhaust from Tokamaks: Problems and Solutions.
\newblock Astrophysics and space science. 1997;256:247-62.

\bibitem{stangeby2000tutorial}
Stangeby P.
\newblock A tutorial on some basic aspects of divertor physics.
\newblock Plasma Physics and Controlled Fusion. 2000;42(12B):B271.

\bibitem{boozer2015stellarator}
Boozer AH.
\newblock Stellarator design.
\newblock Journal of Plasma Physics. 2015;81(6):515810606.

\bibitem{renner2004}
Renner H, Sharma D, Kisslinger J, Boscary J, Grote H, Schneider R.
\newblock Physical aspects and design of the Wendelstein 7-X divertor.
\newblock Fusion science and technology. 2004;46(2):318-26.

\bibitem{ohyabu1992helical}
Ohyabu N, Noda N, Ji H, Akao H, Akaishi K, Ono T, et~al.
\newblock Helical divertor in the large helical device.
\newblock Journal of nuclear materials. 1992;196:276-80.

\bibitem{masuzaki2006overview}
Masuzaki S, Morisaki T, Shoji M, Kubota Y, Watanabe T, Kobayashi M, et~al.
\newblock Overview and future plan of helical divertor study in the Large Helical Device.
\newblock Fusion science and technology. 2006;50(3):361-71.

\bibitem{boozer2018simulation}
Boozer AH, Punjabi A.
\newblock Simulation of stellarator divertors.
\newblock Physics of Plasmas. 2018;25(9).

\bibitem{punjabi2020simulation}
Punjabi A, Boozer AH.
\newblock Simulation of non-resonant stellarator divertor.
\newblock Physics of Plasmas. 2020;27(1).

\bibitem{punjabi2022magnetic}
Punjabi A, Boozer AH.
\newblock Magnetic turnstiles in nonresonant stellarator divertor.
\newblock Physics of Plasmas. 2022;29(1).

\bibitem{boozer2023magnetic}
Boozer AH.
\newblock Magnetic field properties in non-axisymmetric divertors.
\newblock Physics of Plasmas. 2023;30(11).

\bibitem{bader2017hsx}
Bader A, Boozer A, Hegna C, Lazerson S, Schmitt J.
\newblock HSX as an example of a resilient non-resonant divertor.
\newblock Physics of Plasmas. 2017;24(3).

\bibitem{garcia2024resilient}
Garcia K, Bader A, Boeyaert D, Boozer A, Frerichs H, Gerard M, et~al.
\newblock Resilient Stellarator Divertor Characteristics in the Helically Symmetric eXperiment.
\newblock arXiv preprint arXiv:241110611. 2024.

\bibitem{garcia2023exploration}
Garcia K, Bader A, Frerichs H, Hartwell G, Schmitt J, Allen N, et~al.
\newblock Exploration of non-resonant divertor features on the Compact Toroidal Hybrid.
\newblock Nuclear Fusion. 2023;63(12):126043.

\bibitem{allen2024studies}
Allen N, Garcia K, Bader A, Schmitt J, Maurer D, Ennis D, et~al.
\newblock Studies of Non-Resonant Divertor Strike Line Resiliency in the Compact Toroidal Hybrid.
\newblock Bulletin of the American Physical Society. 2024.

\bibitem{strumberger1992magnetic}
Strumberger E.
\newblock Magnetic field line diversion in Helias stellarator configurations: perspectives for divertor operation.
\newblock Nuclear fusion. 1992;32(5):737.

\bibitem{strumberger1992topology}
Strumberger E.
\newblock Topology of divertor field line mapping in HELIAS configurations.
\newblock Contributions to Plasma Physics. 1992;32(3-4):212-8.

\bibitem{nuhrenberg1992development}
N{\"u}hrenberg J, Strumberger E.
\newblock Development of divertor concept for optimized stellarators.
\newblock Contributions to Plasma Physics. 1992;32(3-4):204-11.

\bibitem{parra2024flexible}
Parra F, Baek SG, Churchill M, Demers D, Dudson B, Ferraro N, et~al.
\newblock Flexible Stellarator Physics Facility.
\newblock arXiv preprint arXiv:240704039. 2024.

\bibitem{pablant2024compelling}
Pablant NA, Anderson D, Anderson J, Anderson S, Andruczyk D, Bader A, et~al.
\newblock The Compelling Need for a Mid-Scale Stellarator Facility.
\newblock Los Alamos National Laboratory (LANL), Los Alamos, NM (United States); 2024.

\bibitem{tang2024divertor}
Tang L, Kumar S, Swanson C, Dudt D, Kruger T, Martin M, et~al.
\newblock Divertor design plans for Eos stellarator.
\newblock Bulletin of the American Physical Society. 2024.

\bibitem{poincare1928oeuvres}
Poincar{\'e} H.
\newblock Œuvres de Henri Poincare Tome 1.
\newblock Paris: Gauthier-Villars; 1928.

\bibitem{greene1968two}
Greene JM.
\newblock Two-Dimensional Measure-Preserving Mappings.
\newblock Journal of Mathematical Physics. 1968;9(5):760-8.

\bibitem{arnold1992ordinary_index}
Arnold VI.
\newblock Ordinary differential equations.
\newblock Springer Science \& Business Media; 1992.

\bibitem{solov1970plasma}
Solov’ev L, Shafranov V.
\newblock Plasma confinement in closed magnetic systems.
\newblock In: Reviews of Plasma Physics: Volume 5. Springer; 1970. p. 1-247.

\bibitem{smiet2020bifurcations}
Smiet C, Kramer G, Hudson S.
\newblock Bifurcations of the magnetic axis and the alternating-hyperbolic sawtooth.
\newblock Nuclear Fusion. 2020;60(8):084005.

\bibitem{wei2023invariant}
Wei W, Liang Y.
\newblock Invariant manifold growth formula in cylindrical coordinates and its application for magnetically confined fusion.
\newblock Plasma Science and Technology. 2023;25(9):095105.

\bibitem{wolf1986quantifying}
Wolf A, et~al.
\newblock Quantifying chaos with Lyapunov exponents.
\newblock Chaos. 1986;16:285-317.

\bibitem{kerst1962influence}
Kerst D.
\newblock The influence of errors on plasma-confining magnetic fields.
\newblock Journal of Nuclear Energy Part C, Plasma Physics, Accelerators, Thermonuclear Research. 1962;4(4):253.

\bibitem{duignan2024global}
Duignan N, Perrella D, Pfefferl{\'e} D.
\newblock Global realisation of magnetic fields as $\tfrac{1}{2}$ D Hamiltonian systems.
\newblock arXiv preprint arXiv:240705692. 2024.

\bibitem{meiss1992symplectic}
Meiss J.
\newblock Symplectic maps, variational principles, and transport.
\newblock Reviews of Modern Physics. 1992;64(3):795.

\bibitem{giuliani2024direct}
Giuliani A.
\newblock Direct stellarator coil design using global optimization: application to a comprehensive exploration of quasi-axisymmetric devices.
\newblock Journal of Plasma Physics. 2024;90(3):905900303.

\bibitem{giuliani2024comprehensive}
Giuliani A, Rodr{\'\i}guez E, Spivak M.
\newblock A comprehensive exploration of quasisymmetric stellarators and their coil sets.
\newblock arXiv preprint arXiv:240904826. 2024.

\bibitem{helander2014theory}
Helander P.
\newblock Theory of plasma confinement in non-axisymmetric magnetic fields.
\newblock Reports on Progress in Physics. 2014;77(8):087001.

\bibitem{pyoculus}
;.
\newblock Https://github.com/zhisong/pyoculus.

\end{thebibliography}

\begin{appendix}
\section{Greene's Residue and Lyapunov exponents of fixed points}\label{app:pyoculus}
We use the pyoculus package to find the Jacobian $\MM$ of each fixed point of the QUASR examples shown in section \ref{sec:quasr_db}, and from this the Greene's Residue and the Lyapunov exponent (see section \ref{sec:jacobian}). This data is summarised in table \ref{tab:fixed_points}, with the X-point of the W7-X standard case (with a $5/5$ island chain) included for reference. 

The magnitude of the residue, representing how ``different" a fixed point is from an intact rational surface, is smallest for the three entries corresponding to island chains (the island chain of W7-X and of configuration 74608), and $1-2$ orders of magnitude larger for the other fixed points. It is the largest for the $\iota=0$ fixed point of configuration 1258083. It is interesting to note that the hyperbolic island chain of configuration 1258083 also has a large (negative) residue; the fixed points are strongly perturbed away from an intact surface.

The Lyapunov exponents correlate with Greene's Residue; the greater the magnitude of Greene's Residue, the larger the exponent. It would be interesting to examine the consequences of this on plasma dynamics (heat and particle transport) in the edge.

\begin{table}[h]
\centering
\begin{tabular}{c||ccc} \hline \hline
fixed point & Topological Index & Residue & Lyapunov exponent     \\ \hline 
104183 top & -1 & -1.93 & 2.26   \\  \hline 
104183 bottom & -1 & -1.93 & 2.26   \\  \hline 
74608 edge & -1 & -1.06 & 1.80   \\  \hline 
74608 island X-point & -1 & -0.01 & 0.22   \\  \hline 
74608 island O-point & +1 & 0.03 & n/a   \\  \hline 
1258083 $\imath$ =0 fixed point & -1 & -2.19 & 9.46   \\  \hline 
1258083 $\imath=1$ (a)  & -1 & -1.89 & 2.25   \\  \hline 
1258083 $\imath =1$ (b) & +1 & 1.87 & 1.66   \\  \hline 
W7-X island X-point & -1 & -0.12 & 0.67   \\  \hline 
\end{tabular}
\caption{Properties of fixed points analyzed in the QUASR database and the X-point of the island chain of W7X}
\label{tab:fixed_points}
\end{table}

\end{appendix}
\end{document}